\documentclass[preprint,1p,number]{elsarticle}
\usepackage[usenames,dvipsnames]{color}
\usepackage{graphicx}
\usepackage{amssymb}

\DeclareGraphicsExtensions{.eps}

\newcommand{\secref}[1]{section~\ref{sec_#1}}

\newcommand{\figref}[1]{Fig.~\ref{fig_#1}}
\newcommand{\tblref}[1]{Table~\ref{tbl_#1}}
\newcommand{\eqref}[1]{Eq.~(\ref{eq_#1})}

\journal{Physica A}

\begin{document}

\begin{frontmatter}

\title{Community structure of complex software systems:\\
Analysis and applications}

\author{Lovro \v Subelj\corref{coraut}}
\ead{lovro.subelj@fri.uni-lj.si}
\author{Marko Bajec}
\address{Faculty of Computer and Information Science, University of Ljubljana, Ljubljana, Slovenia}
\cortext[coraut]{Corresponding author}

\begin{abstract}
Due to notable discoveries in the fast evolving field of complex networks, recent research in software engineering has also focused on representing software systems with networks. Previous work has observed that these networks follow scale-free degree distributions and reveal small-world phenomena, when we here explore another property commonly found in complex networks, i.e. community structure. We adopt class dependency networks, where nodes represent software classes and edges represent dependencies among them, and show that these networks reveal significant community structure, characterized by similar properties as observed in other complex networks. However, although intuitive and anticipated by different phenomena, identified communities do not exactly correspond to software packages. We empirically confirm our observations on several networks constructed from \textit{Java} and various third party libraries, and propose different applications of community detection to software engineering.
\end{abstract}


\end{frontmatter}

%%%%%%%%%%%%%%%%%%%%%%%%%%%%%%%%%%%%%%%%%%%%%%%%%%%%%%%%%%%%%%%%%%%%%%%

\section{\label{sec_intro}Introduction}
Analysis of complex real-world networks has led to some significant discoveries in the recent years. Research community has revealed several common properties of various real-world networks~\cite{WS98,BA99,GN02}, including different social, biological, Internet, software and other networks. These properties provide an important insight in the function and structure of general complex networks~\cite{Str01,PDFV05}, moreover, they allow for better comprehension of the underlying real-world systems and thus give prominent grounds for future research in a wide variety of different fields.

In the field of software engineering, network analysis has just recently been adopted to acquire better comprehension of the complex software systems~\cite{Mye03,LHH08,CY09,Koh09}. Nowadays, software represents one of the most diverse and sophisticated human made systems; however, only little is known about the actual structure and quantitative properties of (large) software systems. Cai~and~Yin~\cite{CY09} have denoted this dilemma as \textit{software law problem}, which represents an effort towards identifying and formulating physics-like laws, obeyed by (most) software systems, that could later be applied in practice. However, the main objective of software law problem is in investigating how software looks like.

In the context of employing complex networks analysis, research community has already made several discoveries over the past years. In particular, different authors have observed that networks, constructed from various software systems, follow \textit{scale-free}~\cite{BA99} (i.e. power-law) degree distributions and reveal \textit{small-world}~\cite{WS98} phenomena. We proceed their work by exploring another property commonly found in real-world networks, i.e. \textit{community structure}~\cite{GN02}. The term denotes the occurrence of local structural modules (\textit{communities}) that are groups of nodes densely connected within and only loosely connected with the rest of the network. Communities play crucial roles in many real-world systems~\cite{GD03,PDFV05}, however, the community structure of complex software system networks has not yet been thoroughly investigated. 

Main contributions of our work are as follows. We adopt \textit{class dependency networks}, where nodes represent software classes and edges represent dependencies among them, and show that these networks reveal significant community structure, with similar properties as observed for other complex networks. We also note that network, representing core software library, exhibits less significant community structure. Furthermore, we prove that, although intuitive and anticipated by different phenomena, revealed communities do not (completely) correspond to software packages. Thus, we demonstrate how community detection can be employed to obtain highly modular software packages that still relate to the original packaging.

The rest of the article is structured as follows. First, in~\secref{rw}, we briefly present relevant related work and emphasize the novelty of our research. Next, \secref{cdn} introduces employed class dependency networks. In~\secref{eval} we present empirical evaluation of community structure of class dependency networks, and propose possible applications to software engineering. Last, in~\secref{conc}, we give final conclusions and identify areas of further research.

%%%%%%%%%%%%%%%%%%%%%%%%%%%%%%%%

\section{\label{sec_rw}Related work}
Although software law problem has already been investigated over $30$ years~\cite{Hal77}, research community has only recently begin to employ network analysis to gain better comprehension of the software systems~\cite{LHH08,CY09,Koh09,SB10}. As mentioned above, different authors have observed that networks, constructed from software systems, follow scale-free degree distributions~\cite{VCS02,Mye03,HCK05,CMPS06} and exhibit small-world property~\cite{Mye03,LW04,VS07}. Software networks thus reveal common behavior, similar as observed in other complex networks~\cite{Str01,GN02}. Furthermore, authors have also identified several different phenomena (e.g. software optimization) that might govern such complex behavior~\cite{VCS02,GP04,VS07,ZZLW08}. Moreover, analysis of \textit{clustering}~\cite{WS98} has revealed hierarchical structure in software networks~\cite{Mye03}.

On the other hand, community structure of software networks has not yet been investigated. Several authors have already discussed the notion of communities in the context of software systems~\cite{Mye03,BFNRSVMT06,VS07,JK07,LHH08}, however, no general empirical analysis and formal discussion was ever conducted (due to our knowledge). Still, authors have observed different phenomena that could promote the emergence of community structure in software networks~\cite{BFNRSVMT06,LHH08} and discussed possible applications within software engineering and other sciences~\cite{Mye03,LHH08}.

In a wider context of software networks analysis, different random-walk based measures have been proposed to measure key (i.e. most influential) classes and packages~\cite{ZCDP05,Koh09}. The researchers have also investigated connectedness, robustness, motifs and patterns within software networks~\cite{Mye03,JK07}. Just recently software systems were also treated as evolving complex networks~\cite{CY09}.

%%%%%%%%%%%%%%%%%%%%%%%%%%%%%%%%

\section{\label{sec_cdn}Class dependency networks}
Previous research on the analysis of software systems has employed a variety of different types of software networks (i.e. graphs). In particular, \textit{package}, \textit{class} and \textit{method collaboration graphs}~\cite{Mye03,HCK05}, \textit{subrutine call graphs}~\cite{Mye03}, \textit{software architecture}~\cite{JK07} and \textit{software mirror graphs}~\cite{CY09}, \textit{software architecture maps}~\cite{VCS02}, \textit{inter-package dependency networks}~\cite{LW04} and others~\cite{Mye03,VS07,WKD07}. The networks primarily divide whether they are constructed from source code, byte code or software execution traces, and due to the level of software architecture they represent. However, as discussed in~\secref{rw}, most of these networks share some common characteristics.

%%%%%%%%%%%%%%%

\begin{figure}
\centering
\includegraphics[width=0.50\columnwidth]{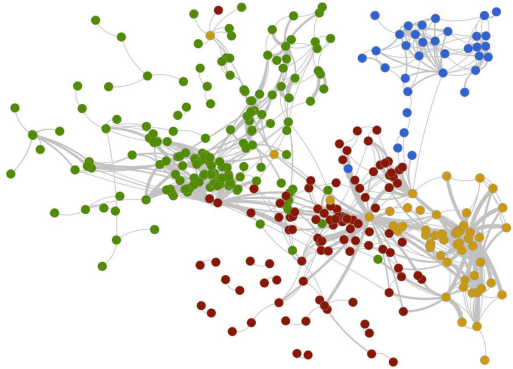}
\caption{\label{fig_jung}Class dependency network for \textit{JUNG} graph and network framework~\cite{OFWSB05}. Node colors indicate four high-level packages of the framework -- \texttt{visualization} (green), \texttt{algorithms} (red), \texttt{graph} (orange) and \texttt{io} (blue). The network reveals rather clear community structure that roughly coincides with the software packages.}
\end{figure}

%%%%%%%%%%%%%%%

For the purpose of this research we introduce \textit{class dependency networks} (\figref{jung}). Here an object-oriented software is represented by an undirected multi-graph $G(N,E)$, where $N$ is the set of nodes and $E$ is the set of edges. Graph $G$ is constructed from software source code in the following manner. Each software class $c$ is represented by a node $n_c\in N$, when edge $\{n_{c_1},n_{c_2}\}\in E$ represents a \textit{dependency} between classes $c_1$ and $c_2$. Dependencies are of four types, namely, \textit{inheritance} (class $c_2$ inherits or implements class $c_1$), \textit{field} ($c_2$ contains a field of type $c_1$), \textit{parameter} ($c_2$ contains a method that takes type $c_1$ as a parameter) and \textit{return} ($c_1$ contains a method that returns type $c_2$).

Note that class dependency networks are constructed merely from the header information of the classes, and their methods and fields. As this information is commonly determined by a group of developers, prior to the actual software development, it is less influenced by the subjective nature of each particular developer. Hence, the networks thus more adequately represent the (intended) structure of some particular software, still, some relevant information might thus be discarded. 

An example of class dependency network is shown in~\figref{jung}. The network reveals strong community structure, furthermore, the communities also roughly coincide with the actual software packages. However, as will be shown in \secref{eval}, modularity of the natural communities, depicted in the network's topology, is much larger than that of the packages, determined by the developers.

%%%%%%%%%%%%%%%%%%%%%%%%%%%%%%%%

\section{\label{sec_eval}Empirical analysis and applications}
In the proceeding sections we present and discuss results of the empirical evaluation of community structure of class dependency networks (\secref{eval_cs}), address the relation between communities and software packages (\secref{eval_sp}) and propose possible applications of community detection to software engineering (\secref{eval_app}). 

The empirical evaluation is done using $8$ class dependency networks constructed\footnotemark[1] from \textit{Java} and several third party libraries (\tblref{net}). The networks range from those with hundreds of nodes to those with several tens of thousands of edges (all isolated nodes have been discarded). Due to generality, networks were selected thus they represent a relatively diverse set of software systems.

\footnotetext[1]{Networks were constructed by parsing \textit{JAR} archives provided by the developers. However, due to various issues, some of the software classes might thus have been discarded.}

%%%%%%%%%%%%%%%

\begin{table}[h]
\centering
\caption{\label{tbl_net}Class dependency networks for different software systems ($|P|$ is the number of packages).}
\begin{tabular}{ccrrr}
\\
\hline\noalign{\smallskip}
Network & Description & $|N|$ & $|E|$ & $|P|$ \\
\noalign{\smallskip}
\hline
\noalign{\smallskip}
\textit{junit} & \textit{JUnit} 4.8.1 (testing framework).~\cite{Aut00junit} & 128 & 470 & 22 \\
\textit{jmail} & \textit{JavaMail} 1.4.3 (mail and messaging framework).~\cite{Aut00jmail} & 220 & 893 & 14 \\
\textit{flamingo} & \textit{Flamingo} 4.1 (GUI component suite).~\cite{Aut00flamingo} & 251 & 846 & 16 \\
\textit{jung} & \textit{JUNG} 2.0.1 (graph and network framework).~\cite{OFWSB05} & 422 & 1730 & 39 \\
\textit{colt} & \textit{Colt} 1.2.0 (scientific computing library).~\cite{Aut00colt} & 520 & 3691 & 19 \\
\textit{org} & \textit{Java} 1.6.0 (\textit{org} namespace).~\cite{Aut00java} & 716 & 7895 & 47 \\
\textit{javax} & \textit{Java} 1.6.0 (\textit{javax} namespace).~\cite{Aut00java} & 2581 & 22370 & 110 \\
\textit{java} & \textit{Java} 1.6.0 (\textit{java} namespace).~\cite{Aut00java} & 2378 & 34858 & 54 \\
\hline
\end{tabular}
\end{table}

%%%%%%%%%%%%%%%

To reveal community structure of each network we employ three community detection algorithms. In particular, a divisive algorithm based on \textit{edge betweenness}~\cite{GN02}, a greedy agglomerative optimization of \textit{modularity} (see below)~\cite{New04,CNM04} and a fast partitional algorithm based on \textit{label propagation}~\cite{RAK07}. The algorithms are denoted \textit{EB}, \textit{MO} and \textit{LP} respectively, whereas, the detailed description is omitted. It should be noted that our objective is not to compare the algorithms, but rather to compare the revealed communities, and thus address their stability.

The community structure, identified by the algorithms, is assessed using \textit{modularity} $Q$~\cite{NG04} that measures the significance of communities due to a selected \textit{null model}. Let $l_i$ be the community (label) of node $n_i\in N$ and let $A_{ij}$ denote the number of edges incident to nodes $n_i, n_j\in N$. Furthermore, let $P_{ij}$ be the expected number of incident edges for $n_i,n_j$ in the null model. The modularity then reads
\begin{eqnarray}
Q & = & \frac{1}{2m}\sum_{n_i,n_j\in N}\left(A_{ij}-P_{ij}\right)\delta(l_i,l_j),
\end{eqnarray}
where $m$ is the number of edges, $m=|E|$, and $\delta$ is the Kronecker delta. The modularity thus measures the fraction of the difference between the number intra-community edges and the expected number of edges in the null model ($Q\in [-1,1]$). Higher values represent stronger community structure. Commonly a random graph with the same degree distribution as the original is selected for the null model. Hence, $P_{ij}=\frac{k_ik_j}{2m}$, where $k_i$ is the degree of node $n_i\in N$. It should be noted that modularity has some known deficiencies, e.g. \textit{resolution limit}~\cite{FB07}, however, it is still widely adopted for the analysis of network community structure.

Furthermore, the identified community structure is also compared to the actual software packages. Let $\mathcal{L}$ be the partition (i.e. communities) revealed by some algorithm and $\mathcal{P}$ the partition that represents software packages (corresponding random variables are $L$ and $P$ respectively). We compare the partitions by computing their \textit{normalized mutual information NMI}~\cite{DDDA05} ($\mbox{\textit{NMI}}\in [0,1]$). Hence, 
\begin{eqnarray}
\mbox{\textit{NMI}} & = & \frac{2I(L,P)}{H(L)+H(P)},
\end{eqnarray}
where $I(L,P)$ is the \textit{mutual information} of the partitions, $I(L,P)=H(L)-H(L|P)$, and $H(L)$, $H(P)$ and $H(L|P)$ are standard and conditional entropies. \textit{NMI} of identical partitions equals $1$, and is $0$ for independent partitions.

%%%%%%%%%%%%%%%%%%%%%%%%%%%%%%%%

\subsection{\label{sec_eval_cs}Community structure of class dependency networks}
Mean modularities obtained with three community detection algorithms for the selected set of class dependency networks (\secref{eval}) can be seen in~\tblref{q}. For all networks except \textit{java}, the algorithms managed to reveal community structures with particularly high values of modularity, i.e. between $0.55$ and $0.75$ on average, where values above $0.30$ are commonly regarded as an indication of (significant) community structure~\cite{RAK07,BGLL08,MHSWL09,LM09b}. The networks thus reveal much stronger community structure than expected in a random network with the same degree distribution. Note also that all of the algorithms obtain high modularities for all of the networks considered. This indicates rather stable communities, strongly depicted in the networks' topologies.

%%%%%%%%%%%%%%%

\begin{table}
\centering
\caption{\label{tbl_q}Mean modularities $Q$ obtained for class dependency networks of different software systems. Values were computed from $100$ iterations ($10$ iterations for \textit{EB} algorithm), where missing values could not be recovered due to limited time resources. Modularities of the natural community structures, depicted in the networks' topologies (i.e. extracted by the algorithms), are much larger than those of the actual software packages.}
\begin{tabular}{cccccc}
\\
\hline\noalign{\smallskip}
Network & \textit{EB} & \textit{MO} & \textit{LP} & $P^+$ & $P$ \\
\noalign{\smallskip}
\hline
\noalign{\smallskip}
\textit{junit} & 0.5587 & 0.5759 & 0.5542 & 0.1140 & 0.0893 \\
\textit{jmail}  & 0.5607 & 0.5972 & 0.5401 & 0.2350 & 0.2086 \\
\textit{flamingo} & 0.6466 & 0.6823 & 0.6485 & 0.2870 & 0.2511 \\
\textit{jung} & 0.7210 & 0.7324 & 0.6874 & 0.3279 & 0.3212 \\
\textit{colt} & - & 0.6025 & 0.5599 & -0.0158 & -0.0332 \\
\textit{org} & - & 0.5599 & 0.5254 & 0.1847 & 0.1830 \\
\textit{javax} & - & 0.7667 & 0.7422 & 0.3119 & 0.2907 \\
\textit{java} & - & 0.4664 & 0.4132 & 0.2269 & 0.2206 \\
\hline
\end{tabular}
\end{table}

%%%%%%%%%%%%%%%

In the case of \textit{java} network, observe that the average degree is considerably larger than for other networks (\tblref{net}). Hence, the network is extremely dense and the communities are thus only loosely defined. Consequently the algorithms fail to attain any significant community structure; however, as the network represents the core of \textit{Java} programming language, it is expected to convey less modular structure.

\begin{figure}[b]
\centering
\includegraphics[width=1.00\columnwidth]{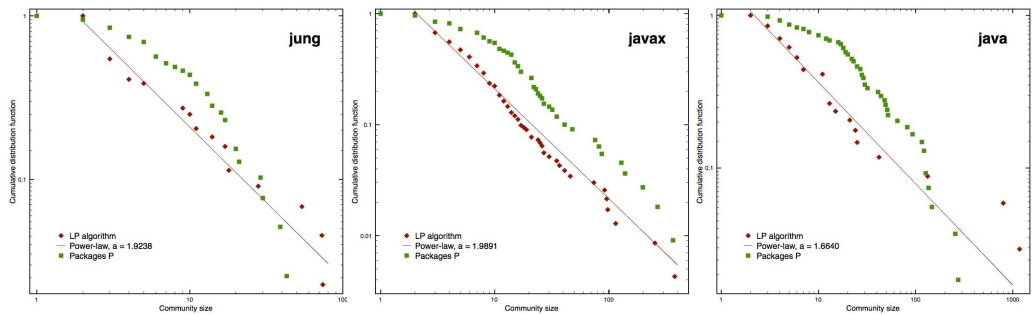}
\caption{\label{fig_sizes}Cumulative distribution functions of community sizes for \textit{jung}, \textit{javax} and \textit{java} networks. The distributions revealed by \textit{LP} algorithm (roughly) follow power-laws with the exponents $\alpha$ shown (i.e. $P(s)\sim s^{-\alpha}$, where $s$ is the community size); however, the distributions of package sizes are not characterized by power-laws (e.g. log-normal distributions).}
\end{figure}

In~\figref{sizes} we show the (cumulative) distributions of community sizes obtained with \textit{LP} algorithm for \textit{jung}, \textit{javax} and \textit{java} networks. Interestingly, the distributions (roughly) follow power-laws with the exponents $\alpha$ around $2$ (i.e. $P(s)\sim s^{-\alpha}$, where $s$ is the community size). Scale-free distribution of community sizes is a common property, observed also in other complex networks~\cite{New04,PDFV05}; furthermore, the values of $\alpha$ also coincide with values found for other networks, where authors commonly report $\alpha$ between $1$ and $3$~\cite{CNM04,New04,RCCLP04,PDFV05}.

%%%%%%%%%%%%%%%

%%%%%%%%%%%%%%%

We conclude that class dependency networks contain significant community structure that also reveals similar properties as observed in other complex networks. Thus, besides scale-free degree distributions and small-world effect, software networks reflect another common network phenomena, i.e. community structure.

To further address the validity of our results, we briefly discuss different phenomena that could promote the emergence of community structure in software networks. Li~et~al.~\cite{LHH08} and Jenkins~and~Kirk~\cite{JK07} have discussed the influence of internal \textit{cohesion}, i.e. functional strength of the components, and external \textit{coupling}, i.e. inter-dependencies among components, on the structure of software systems (and networks). Highly modular software should clearly demonstrate \textit{minimum coupling-maximum cohesion} principle~\cite{SMC99}, which would naturally promote the emergence of strong structural modules within software networks. The modularity of software networks thus reflects the modularity of underlying software systems.

Furthermore, Baxter~et~al.~\cite{BFNRSVMT06} have emphasized that object-oriented software is commonly developed according to \textit{Lego hypothesis}~\cite{Szy98}. The hypothesis states that software is constructed out of a larger number of smaller components that are relatively independent of each other. This phenomena should clearly reflect in software networks, where components should emerge as network communities.

\begin{figure}
\centering
\includegraphics[width=1.00\columnwidth]{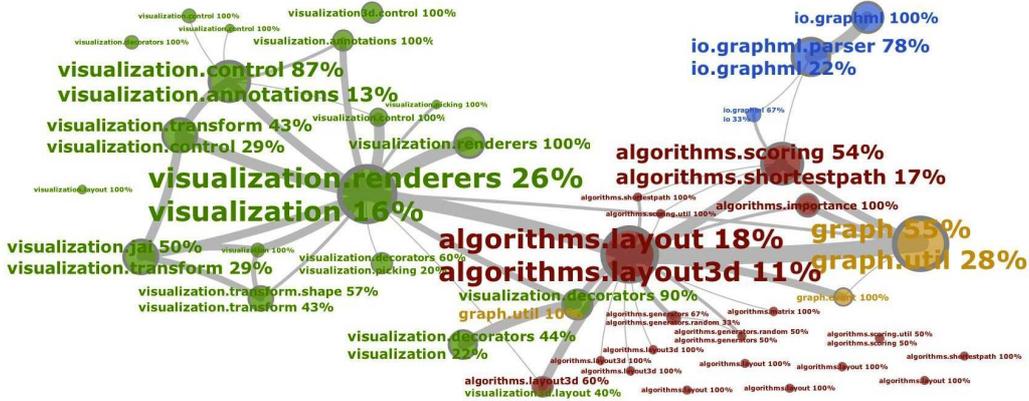}
\caption{\label{fig_cjung}Community network for \textit{jung} class dependency network (\figref{jung}) revealed by \textit{LP} algorithm (modularity equals $Q=0.7062$). For each community we show the distribution of classes over software packages (weakly represented packages are not shown), where colors indicate four high-level packages of the framework (see~\figref{jung}). Communities clearly distinguish between high-level packages, but they do not completely coincide with the actual (bottom-most) packages.} 
\end{figure}

In summary, software networks enclose a strong natural tendency to form community structure. In the case of class dependency networks, communities should, due to the above discussion and by intuition, correspond to software packages (\figref{cjung}). This aspect is thoroughly explored and discussed in the proceeding section.

%%%%%%%%%%%%%%%

%%%%%%%%%%%%%%%

%%%%%%%%%%%%%%%%%%%%%%%%%%%%%%%%

\subsection{\label{sec_eval_sp}Relation of network communities to software packages}
The analysis of the relation between network communities and software packages reveals that packages are considerably different than communities. We first note that packages do not feature \textit{connectedness} in class dependency networks (exact results are omitted). The latter is regarded as a basic property of communities and states that communities should correspond to connected sets of nodes. As a consequence, software packages can comprise of disconnected sets of nodes, which is an indicator of relatively poor modular structure. 

Let $P$ represent the actual (bottom-most) software packages and let $P^+$ represent packages that feature connectedness (i.e. disconnected packages are treated as several different packages). \tblref{q} shows modularities of software packages for the analyzed class dependency networks. The values are considerably lower than modularities of the natural community structures, revealed in the networks' topologies (i.e. extracted by the algorithms), and cannot be regarded as an indication of significant modular structure. Moreover, in~\figref{sizes} we show the distributions of package sizes for \textit{jung}, \textit{javax} and \textit{java} networks. The distributions are obviously not characterized by power-laws, as observed in the case of communities (distributions are, e.g., log-normal or stretched exponential, which coincides with the observations in~\cite{BFNRSVMT06}). Last, we also (directly) compare the packages with network communities by computing \textit{NMI} of the corresponding partitions (\tblref{nmi}). The results further confirm above observations -- software packages only weakly relate to network communities and are not characterized by the same laws or properties.

%%%%%%%%%%%%%%%

\begin{table}
\centering
\caption{\label{tbl_nmi}Peak (maximum) \textit{NMI} between network communities, extracted by the algorithms, and software packages $P$ for different class dependency networks. Values were computed from $100$ iterations ($10$ iterations for \textit{EB} algorithm). The results indicate relatively poor correspondence between natural network communities and software packages.}
\begin{tabular}{ccccc}
\\
\hline\noalign{\smallskip}
Network & \textit{EB} & \textit{MO} & \textit{LP} & $P^+$ \\
\noalign{\smallskip}
\hline
\noalign{\smallskip}
\textit{junit} & 0.6605 & 0.5823 & 0.6285 & 0.8412 \\
\textit{jmail} & 0.5300 & 0.5248 & 0.5553 & 0.8379 \\
\textit{flamingo} & 0.5686 & 0.5408 & 0.5590 & 0.7882 \\
\textit{jung} & 0.6011 & 0.6094 & 0.6887 & 0.9187 \\ 
\textit{colt} & - & 0.4784 & 0.5277 & 0.6507 \\
\textit{org} & - & 0.5301 & 0.5385 & 0.9123 \\
\textit{javax} & - & 0.6365 & 0.6826 & 0.8096 \\
\textit{java} & - & 0.3453 & 0.3063 & 0.8386 \\
\hline
\end{tabular}
\end{table}

%%%%%%%%%%%%%%%

We stress that the origin of the disparity between network communities and software packages is not entirely evident. The lack of connectedness of software packages, and low values of modularity, suggest that class dependency networks give poor representation of software systems or disregard some relevant relations among classes (form the perspective of software packages). However, different distributions of sizes clearly show that there is some additional departure between the communities and software packages, which is independent of the actual network representation (i.e. class dependencies).

Last, we discuss a particularly low value of modularity for \textit{colt} library packages (\tblref{q}). As the library represents a core framework for scientific computing, where the performance is often of greater importance than extensibility, maintenance and modular structure, it is expected for the system to exhibit only poor modular structure. The modularity of software packages thus reflects the modularity of the underlying software system, which in fact motivates the application, presented in the proceeding section.

%%%%%%%%%%%%%%%%%%%%%%%%%%%%%%%%

\subsection{\label{sec_eval_app}Applications of community detection to software engineering}
\begin{figure}
\centering
\includegraphics[width=1.00\columnwidth]{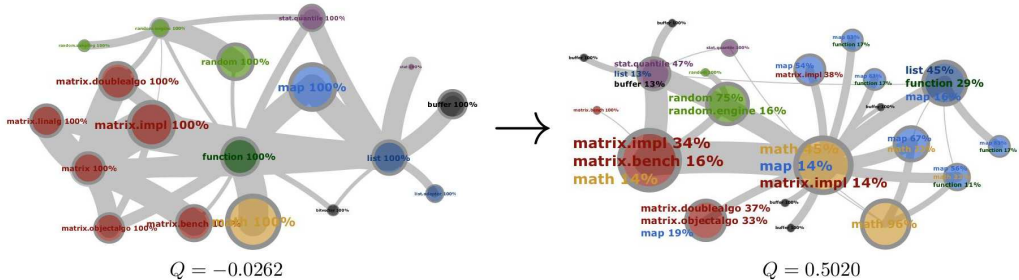}
\caption{\label{fig_colt}Community networks for class dependency network, representing classes within \texttt{cern.colt} and \texttt{cern.jet} packages of \textit{colt} library (reduced to the largest connected component). Networks correspond to the original software packages $P$ (left) and communities, revealed with \textit{LP} algorithm by refining software packages $P$ (right). For each community we show the distribution of classes over software packages, where colors indicate high-level packages of the framework. Refined communities (i.e. packages) obtain significantly higher modularity and can still be related to the original packaging.} 
\end{figure}

Due to weak modular structure of software packages, an obvious application of community detection to software engineering is to reveal highly modular packaging of software systems (\figref{colt}). The choice of class dependencies (i.e. type of the network) is in that case of course arbitrary. However, simply applying a community detection algorithm to employed networks would often prove useless, as the identified communities would only hardly be mapped to the existing software packages. The latter is vital due to the comprehension of the results. A simple solution is to start with the communities that represent original software packages, and refine them, using some community detection algorithm. The algorithm should thus merely refine and merge the communities, where no new communities (i.e. labels) should be introduced. This preserves original software packages, their hierarchy and identifiers, which enables complete comprehension of the final results. An example can be seen in~\figref{colt}.

%%%%%%%%%%%%%%%

%%%%%%%%%%%%%%%

\begin{figure}
\centering
\includegraphics[width=1.00\columnwidth]{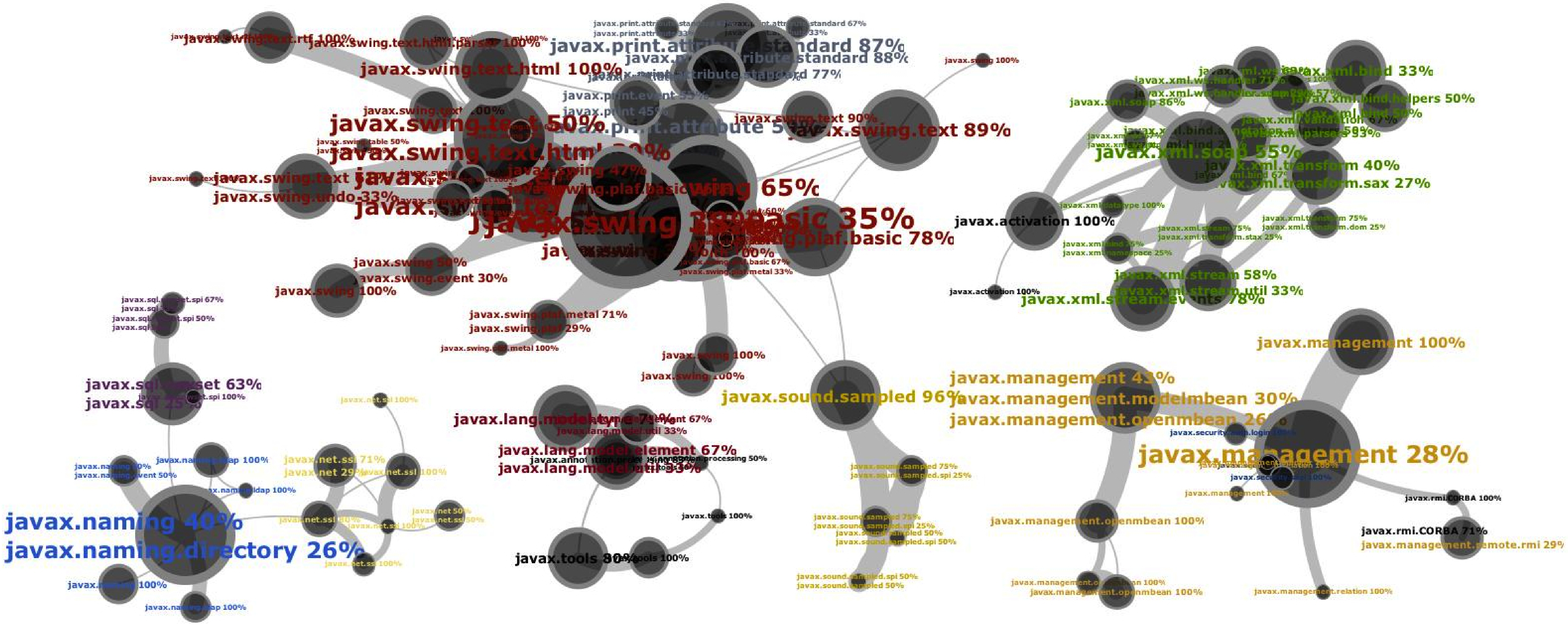}
\caption{\label{fig_cjavax}Community network for \textit{javax} class dependency network revealed by \textit{LP} algorithm (only the largest five connected components are shown; modularity equals $Q=0.7318$). For each community we show the distribution of classes over software packages, where colors indicate high-level packages of the framework. The representation gives a clear insight into the structure of the \textit{javax} namespace, and shows relations (i.e. dependencies) among different sub-packages of the system.}
\end{figure}

Another obvious application to software engineering is (network) \textit{abstraction}. Community detection can be employed to obtain a clear representation of software systems on a relatively high level of abstraction. Furthermore, one can also address the \textit{centrality}~\cite{Fre77,Fre79} (or other measures of influence) of the identified communities, to expose key nodes and structures throughout the entire system~\cite{ZCDP05,Koh09}. A simple application of community detection to software abstraction can be seen in~\figref{cjavax} (and~\figref{cjung}).

The article represents seminal work in the area of applying network community detection methods and techniques in software engineering. However, further work is needed to design sophisticated applications that would be of considerable benefit in practice.

%%%%%%%%%%%%%%%

%%%%%%%%%%%%%%%

%%%%%%%%%%%%%%%%%%%%%%%%%%%%%%%%

\section{\label{sec_conc}Conclusion}
The article explores community structure of networks, constructed from complex software systems (i.e. class dependency networks). The main contribution is in showing that software networks reveal significant community structure, characterized by similar properties as commonly observed for other complex networks. Software networks thus reveal another general network phenomena, besides scale-free degree distributions and small-world effect, which is a prominent observation for the software-law problem. Furthermore, the results are of even greater importance, as software represents one of the most complex human made systems.

Future work will mainly focus on considering other types of class dependency networks that will include additional relations among classes. Moreover, we will introduce the notions of \textit{positive} and \textit{negative} relations, to more adequately model similarity and diversity among software classes. The main objective will be to establish further understanding of (community) structure of class dependency networks, and to assess its relation to software packages. The results could thus promote various novel applications in the software engineering domain.

%%%%%%%%%%%%%%%%%%%%%%%%%%%%%%%%%%%%%%%%%%%%%%%%%%%%%%%%%%%%%%%%%%%%%%%


\begin{thebibliography}{10}
\expandafter\ifx\csname url\endcsname\relax
  \def\url#1{\texttt{#1}}\fi
\expandafter\ifx\csname urlprefix\endcsname\relax\def\urlprefix{URL }\fi
\expandafter\ifx\csname href\endcsname\relax
  \def\href#1#2{#2} \def\path#1{#1}\fi

\bibitem{WS98}
D.~J. Watts, S.~H. Strogatz, Collective dynamics of 'small-world' networks,
  Nature 393~(6684) (1998) 440--442.

\bibitem{BA99}
A.~Barab\'{a}si, R.~Albert, Emergence of scaling in random networks, Science
  286~(5439) (1999) 509--512.

\bibitem{GN02}
M.~Girvan, M.~E.~J. Newman, Community structure in social and biological
  networks, in: Proceedings of the National Academy of Sciences of United
  States of America, 2002, pp. 7821--7826.

\bibitem{Str01}
S.~H. Strogatz, Exploring complex networks, Nature 410 (2001) 268.

\bibitem{PDFV05}
G.~Palla, I.~Der\'{e}nyi, I.~Farkas, T.~Vicsek, Uncovering the overlapping
  community structure of complex networks in nature and society, Nature 435
  (2005) 814.

\bibitem{Mye03}
C.~R. Myers, Software systems as complex networks: Structure, function, and
  evolvability of software collaboration graphs, Phys. Rev. E 68~(2).

\bibitem{LHH08}
D.~Li, Y.~Han, J.~Hu, Complex network thinking in software engineering, in:
  Proceedings of the International Conference on Computer Science and Software
  Engineering, 2008, pp. 264--268.

\bibitem{CY09}
K.~Cai, B.~Yin, Software execution processes as an evolving complex network,
  Inform. Sciences 179~(12) (2009) 1903--1928.

\bibitem{Koh09}
G.~A. Kohring, Complex dependencies in large software systems, Adv. Complex
  Syst. 12~(6) (2009) 565--581.

\bibitem{GD03}
P.~Gleiser, L.~Danon, Community structure in jazz, Adv. Complex Syst. 6~(4)
  (2003) 565.

\bibitem{Hal77}
M.~H. Halstead, Elements of software science, Elsevier, 1977.

\bibitem{SB10}
L.~Subelj, M.~Bajec, Unfolding network communities by combining defensive and
  offensive label propagation, in: Proceedings of the International Workshop on
  the Analysis of Complex Networks, 2010, pp. 87--104.

\bibitem{VCS02}
S.~Valverde, R.~F. Cancho, R.~V. Sole, Scale-free networks from optimal design,
  Europhys. Lett. 60~(4) (2002) 512.

\bibitem{HCK05}
D.~{Hyland-Wood}, D.~Carrington, S.~Kaplan, Scale-free nature of java software
  package, class and method collaboration graphs, Proceedings of the
  International Symposium on Empirical Software Engineering.

\bibitem{CMPS06}
G.~Concas, M.~Marchesi, S.~Pinna, N.~Serra, On the suitability of yule process
  to stochastically model some properties of object-oriented systems, Physica A
  370~(2) (2006) 817--831.

\bibitem{LW04}
N.~{LaBelle}, E.~Wallingford, Inter-package dependency networks in open-source
  software, e-print {arXiv:0411096v1}.

\bibitem{VS07}
S.~Valverde, R.~V. Sole, Hierarchical small worlds in software architecture,
  Dynam. Cont. Dis. Ser. B 14 (2007) 1--11.

\bibitem{GP04}
A.~A. Gorshenev, Y.~M. Pis'mak, Punctuated equilibrium in software evolution,
  Phys. Rev. E 70~(6) (2004) 1--4.

\bibitem{ZZLW08}
X.~Zheng, D.~Zeng, H.~Li, F.~Wang, Analyzing open-source software systems as
  complex networks, Physica A 387~(24) (2008) 6190--6200.

\bibitem{BFNRSVMT06}
G.~Baxter, M.~Frean, J.~Noble, M.~Rickerby, H.~Smith, M.~Visser, H.~Melton,
  E.~Tempero, Understanding the shape of java software, in: Proceedings of the
  {ACM} {SIGPLAN} International Conference on {Object-Oriented} Programming,
  Systems, Languages, and Applications, 2006, pp. 397--412.

\bibitem{JK07}
S.~Jenkins, S.~Kirk, Software architecture graphs as complex networks: A novel
  partitioning scheme to measure stability and evolution, Inform. Sciences
  177~(12) (2007) 2587--2601.

\bibitem{ZCDP05}
A.~Zaidman, T.~Calders, S.~Demeyer, J.~Paredaens, Applying webmining techniques
  to execution traces to support the program comprehension process, in:
  Proceedings of the European Conference on Software Maintenance and
  Reengineering, 2005, pp. 134--142.

\bibitem{WKD07}
L.~Wen, D.~Kirk, R.~G. Dromey, Software systems as complex networks, in:
  Proceedings of the {IEEE} International Conference on Cognitive Informatics,
  2007, pp. 106--115.

\bibitem{OFWSB05}
J.~{O'Madadhain}, D.~Fisher, S.~White, P.~Smyth, Y.~biao Boey, Analysis and
  visualization of network data using {JUNG}, J. Stat. Softw. 10~(2) (2005)
  1--35.

\bibitem{Aut00junit}
{JUnit} testing framework, http://junit.org/.

\bibitem{Aut00jmail}
{JavaMail} mail and messaging framework,
  http://java.sun.com/products/javamail/.

\bibitem{Aut00flamingo}
Flamingo {GUI} component suite, https://flamingo.dev.java.net/.

\bibitem{Aut00colt}
Colt scientific computing library, http://acs.lbl.gov/software/colt/.

\bibitem{Aut00java}
Java language, http://java.sun.com/.

\bibitem{New04}
M.~E.~J. Newman, Detecting community structure in networks, Eur. Phys. J. B
  38~(2) (2004) 321--330.

\bibitem{CNM04}
A.~Clauset, M.~E.~J. Newman, C.~Moore, Finding community structure in very
  large networks, Phys. Rev. E 70~(6) (2004) 066111.

\bibitem{RAK07}
U.~N. Raghavan, R.~Albert, S.~Kumara, Near linear time algorithm to detect
  community structures in large-scale networks, Phys. Rev. E 76~(3) (2007)
  036106.

\bibitem{NG04}
M.~E.~J. Newman, M.~Girvan, Finding and evaluating community structure in
  networks, Phys. Rev. E 69~(2) (2004) 026113.

\bibitem{FB07}
S.~Fortunato, M.~Barthelemy, Resolution limit in community detection, in:
  Proceedings of the National Academy of Sciences of United States of America,
  2007, pp. 36--41.

\bibitem{DDDA05}
L.~Danon, A.~{D\'{i}az-Guilera}, J.~Duch, A.~Arenas, Comparing community
  structure identification, J. Stat. Mech. P09008.

\bibitem{BGLL08}
V.~D. Blondel, J.~Guillaume, R.~Lambiotte, E.~Lefebvre, Fast unfolding of
  communities in large networks, J. Stat. Mech. P10008.

\bibitem{MHSWL09}
J.~Mei, S.~He, G.~Shi, Z.~Wang, W.~Li, Revealing network communities through
  modularity maximization by a contraction{\textendash}dilation method, New J.
  Phys. 11~(4).

\bibitem{LM09b}
X.~Liu, T.~Murata, Advanced modularity-specialized label propagation algorithm
  for detecting communities in networks, Physica A 389~(7) (2009) 1493.

\bibitem{RCCLP04}
F.~Radicchi, C.~Castellano, F.~Cecconi, V.~Loreto, D.~Parisi, Defining and
  identifying communities in networks, in: Proceedings of the National Academy
  of Sciences of the United States of America, Vol. 101, 2004, pp. 2658--2663.

\bibitem{SMC99}
W.~P. Stevens, G.~J. Myers, L.~L. Constantive, Structured design, {IBM} Syst.
  J. 38~(2) (1999) 231--256.

\bibitem{Szy98}
C.~Szyperski, Component software: Beyond object-oriented programming,
  {Addison-Wesley}, 1998.

\bibitem{Fre77}
L.~Freeman, A set of measures of centrality based on betweenness, Sociometry
  40~(1) (1977) 35--41.

\bibitem{Fre79}
L.~C. Freeman, Centrality in social networks: Conceptual clarification, Soc.
  Networks 1~(3) (1979) 215--239.

\end{thebibliography}
\end{document}